    \documentclass[12pt,psf,epsf]{article}
\usepackage[centertags]{amsmath}
\usepackage{amssymb}
\usepackage{graphicx}
\usepackage{epsfig}
\usepackage{ulem}
\usepackage[english]{babel}
\usepackage{array}
\usepackage{amsthm}
\usepackage{latexsym}
\usepackage[mathcal]{euscript}
\pdfoutput=1
\usepackage{epsfig}

 \usepackage{jheppub}
 \usepackage{hyperref}

\usepackage{subfigure}
\usepackage{psfrag}

\usepackage{epsfig}
\usepackage[latin1]{inputenc}
\usepackage{float}
\usepackage{graphicx}
\usepackage{cancel}
\usepackage{mathrsfs}
\usepackage{amssymb}
\usepackage{amsfonts}
\usepackage{amsmath}
\usepackage{slashed}
\usepackage{graphicx}
\usepackage{bm}
\usepackage{color}
\newcommand{\be}{\begin{equation}}
\newcommand{\ee}{\end{equation}}
\newcommand{\bea}{\begin{eqnarray}}
\newcommand{\eea}{\end{eqnarray}}
\newcommand{\bwt}{\begin{widetext}}
\newcommand{\ewt}{\end{widetext}}

\newcommand{\bi}{\begin{itemize}}
\newcommand{\ei}{\end{itemize}}

\newcommand\cC{{\cal C}}

\usepackage{setspace}

\begin{document}

\title {Holographic complexity of local quench at finite temperature}

\author{Dmitry S. Ageev}
\affiliation{Steklov Mathematical Institute, Russian Academy of Sciences, Gubkin str. 8, 119991
Moscow, Russia}

\emailAdd{ageev@mi-ras.ru}

\abstract{This paper is devoted to the study of the evolution of holographic complexity  after a local perturbation of the system at finite temperature.  We calculate  the complexity using both the complexity=action(CA) and the complexity=volume(CA) conjectures and  find that the CV complexity of the total state shows the unbounded late time linear growth. The CA computation shows linear growth with fast saturation to a constant value. We estimate the CV and CA complexity linear growth coefficients and show, that finite temperature  leads to violation of the Lloyds bound for CA complexity. Also it is shown that for composite system after the local quench the state with minimal entanglement may correspond to the maximal complexity.  }

\maketitle

\newpage

\section{Introduction}

The recent progress in understanding of  the relation between the gravitational degrees of freedom and quantum phenomena in the AdS/CFT correspondence include the  holographic entanglement entropy \cite{Ryu:2006bv}-\cite{Swingle:2009bg}, the scrambling, black hole information paradox and the quantum chaos \cite{Susskind:2014rva}-\cite{Susskind:2018pmk}. Recently the notion of the quantum complexity of the state and holographic complexity have attracted considerable attention. There are different holographic proposals and conjectures concerning the quantum complexity  \cite{Stanford:2014jda}-\cite{Abt:2017pmf}.   The CV("complexity=volume") conjecture relates complexity of a state to volume of a hypersurface in the bulk in the constant otime slice.  The CA("complexity=action") conjecture states that the complexity equals to the on-shell gravitational action evaluated in some special region of the bulk. Also there are different covariant generalizations of these proposals \cite{Susskind:2014rva}-\cite{Abt:2017pmf}. There are different setups where the holographic complexity of different proposals has been investigated. The first setup that has been considered is the  thermofield double state evolution in \cite{Susskind:2014rva}, with the generalizations in \cite{Brown:2015bva},\cite{Brown:2015lvg},\cite{Carmi:2017jqz}-\cite{Alishahiha:2018tep}. Different complexity and entanglement proposals have been investigated for the state evolving after the global perturbation \cite{Alishahiha:2018tep},\cite{Moosa:2017yiz}-\cite{Fan:2018xwf}. 

The process when the system evolves after the local perturbation is called the local quench. The two-dimensional CFT  admits well defined description of the local quenches \cite{Calabrese:2007mtj}. The local quench has a well-defined gravity dual \cite{Nozaki:2013wia} which has been investigated in the different versions further in \cite{Asplund:2014coa}-\cite{DeJonckheere:2018pbi}. In  \cite{Ageev-quench,Ageev-quench-2} we have investigated different holographic complexity proposals  in the system following the local perturbation. We have considered the $AdS_3$ Poincare patch deformed by the static particle  \cite{Nozaki:2013wia} as the background dual to the local quench. We have shown that the CV and  CA prescriptions lead to a  qualitatively different behaviour for a subsystems and compared them to the evolution of the entanglement entropy. 
The entanglement and the complexity  show  quite similar behaviour in this process. Also we have shown, that the CA complexity for a local perturbation at the initial time moment saturates Lloyds bound on the complexification rate.

This paper extends this study for the volume and the action complexities  to the 2d CFT states at finite temperature evolving after the local perturbation. The holographic dual of this process is given by the one-sided planar BTZ black hole perturbed by the particle falling on the horizon \cite{Caputa:2014eta}. The entanglement of the semi-infinite subsystem after the finite temperature local quench  does not show the late-time unbounded growth in contrast to  $T=0$ case. Instead of the late-time growth one can show that the entanglement saturates to some constant value. We show that even after the time of the entanglement equilibration the CV complexity  shows unbounded linear growth. This is similar to the picture observed in the holographic description of complexity in the thermofield double state: the complexity growth take place  even after the system is scrambled \cite{Stanford:2014jda}. 
We estimate the linear growth coefficients for both CA and CV conjectures. 
The CA conjecture result (in the probe approximation) gives us, that at the finite temperature the Lloyd bound is violated for all values of temperature. Also the straightforward estimation of the CA complexity does not lead us to the unbounded linear late time growth. 
The CV estimation gives that this bound is violated for some region of parameters of the system like temperature $T$ or perturbing operator dimension $h$.
In the end of the paper we compare the evolution aspects of entanglement and the complexity of the composite system (two disjoint intervals). We find, that the  CV complexity estimation for such system shows (local) minimum of the entanglement at points where the complexity is maximal. In comparison to $T=0$ case the complexity growth is much faster.

The paper is organized in the following way:

In Sec.\ref{sec:setup}, we describe the setup of the holographic local quench.

Sec.\ref{sec:ent} is devoted to the derivation of the entanglement entropy evolution of semi-infinite subsystem evolution.

Sec.\ref{sec:CA} and Sec.\ref{sec:CV} are devoted to  different approximations to the CA and the CV complexity and their computation.

Finally Sec.\ref{sec:disc} is devoted to the discussion.

\section{Holographic description of CFT local quench  at finite temperature  }\label{sec:setup}
The process when the localized perturbation is excited at the initial  time moment and then subsequently evolves  is called local quench.  There are different protocols of local quench in the CFT. In this paper we restrict ourselves to a 2d CFT and assume that the quench is given by the insertion of the primary  operator with the scaling dimension $h$ at the point $x=0$ at time $t=0$. The holographic dual of this quench is given by the particle of fixed energy $E$  deforming the background metric and this corresponds to localized  excitation carrying the same energy $E$ in the CFT. In the case of $2d$ CFT one can obtain the analytical expression for the metric dual to the finite temperature local quench \cite{Caputa:2014eta}. Let us consider the one-sided planar BTZ black hole with the metric
  \bea\label{pbtz}
&&ds^2=\frac{L^2}{z^2}(-f(z)dt^2+\frac{dz^2}{f(z)}+dx^2),\\
&&f(z)=1-Mz^2,
 \eea
with temperature
\be 
T=\frac{1}{2 \pi z_h},\,\,\,\,M=1/z_h^2.
\ee
The particle is falling on the black hole horizon $z=z_h$ following  the trajectory $z=z(t)$  and deforming the vacuum  metric \eqref{pbtz}. The deformed metric  satisfies the  equations of motion
\be\label{eqR}
R^{\mu\nu}-\frac{1}{2}g^{\mu\nu}R^{\mu\nu}+\Lambda g^{\mu\nu}=T^{\mu\nu}, 
\ee 
where $T^{\mu\nu}$ is the stress energy tensor  for the static particle 
\be 
T^{\mu\nu}=\frac{8\pi mG_N}{\sqrt{-g}}\cdot \frac{\partial_tX^\mu\partial_tX^\nu}
{\sqrt{-g_{\mu\nu}\cdot \partial_t  X^\mu(t)\cdot \partial_t X^\nu(t)}}\cdot \delta(z-z(t))\cdot \delta^{d-1}(x_i),
\ee 
where the explicit form of the particle trajectory is given by
\be\label{eq:traj}
z(t)=\frac{\beta}{2\pi}\sqrt{1-\left(1-(\frac{2\pi\varepsilon}{\beta})^2\right)\Big(1-\tanh^2(\frac{2\pi}{\beta}t)\Big)},\,\,\,\,\beta=\frac{1}{T}.
\ee
Parameter $\varepsilon$  is given by $z(0)=\varepsilon$ and in the dual language corresponds to some "smearing" of operator that perturbs the system.
The particle action has the form 
\be 
S=-mL  \int \frac{1}{z} \sqrt{f(z)-\frac{\dot z^2}{f(z)}}dt,
\ee 
where $m$ is the mass of  particle and the dot  denotes the derivative with respect to time $t$. The conformal dimension $h$ of the corresponding quench operator in the dual CFT is given by relation 
\bea\label{m1}
 h =\frac{m L}{2},
\eea
and the particle has the energy 
\be 
E=mL\frac{\sqrt{1-4\pi^2 T^2 \varepsilon^2}}{\varepsilon}.
\ee 
Instead of straightforward solution of equations \eqref{eqR} it is convenient to use the following trick to obtain the explicit form of the dual metri Consider the global $AdS_3$ deformed by the static particle
\be \label{metr1}
ds^2=-d\tau^2
   \left(L^2-\mu+R^2\right)+R^2 d\phi^2+\frac{L^2 dR^2 }{L^2-\mu+R^2},
\ee
where the particle rests at $R=0$. Depending on  parameter $\mu=8GmL^2$ this space corresponds to the conical defect for $\mu<L^2$, and the BTZ black hole for $\mu>L^2$. 
One can relate the global $AdS$ and the coordinates of the BTZ black hole \eqref{pbtz} as follows
\bea\label{coordtr}
&&\phi =\arctan \left(\frac{\varepsilon \sqrt{M} \sinh \left(\sqrt{M}
   x\right)}{\sqrt{1-M z^2} \cosh \left(\sqrt{M}
   t\right)-\sqrt{1-\varepsilon^2 M} \cosh \left(\sqrt{M}
   x\right)}\right),\\ \nonumber
&&\tau = -\arctan\left(\frac{\varepsilon \sqrt{M} \sqrt{1-M z^2} \sinh
   \left(\sqrt{M} t\right)}{\sqrt{1-\varepsilon^2 M} \sqrt{1-M z^2}
   \cosh \left(\sqrt{M} t\right)-\cosh \left(\sqrt{M}
   x\right)}\right),\\ \nonumber
&&R=\frac{L }{2 \varepsilon M z}\sqrt{{\cal A}_1+{\cal A}_2},\\ \nonumber
  &&{\cal A}_1=-8 \sqrt{1-\varepsilon^2 M} \sqrt{1-M z^2}
   \cosh \left(\sqrt{M} t\right) \cosh \left(\sqrt{M} x\right)-4
   \varepsilon^2 M+3-M z^2,\\ \nonumber
   &&{\cal A}_2=\left(2-2 M z^2\right) \cosh ^2\left(\sqrt{M}
   t\right)+\left(1-M z^2\right) \cosh \left(2 \sqrt{M} t\right)+2
   \cosh \left(2 \sqrt{M} x\right).
\eea
Map \eqref{coordtr} relates the static particle in the patch \eqref{metr1} at $R=0$ and the trajectory \eqref{eq:traj} of the  particle falling on the BTZ black hole horizon. Thus applying coordinate transformation \eqref{coordtr} we obtain the full backreacted solution of \eqref{eqR}. The solution is time-dependent non-diagonal metric $g$ of complicated, however explicit form
\be 
ds^2=g_{ij}dx^idx^j
\ee
where $x_i$ corresponds to coordinates $t$ and $x$. We do not write metric $g_{ij}$ down explicitly in the text because it has very complicated form.  In what follows we also denote the dual metric as $g$.

\section{Entanglement dynamics after local quench}\label{sec:ent}
First we describe  the  entanglement entropy evolution in the system dual to the holographic setup described in Sec.\ref{sec:setup}. The entanglement evolution in this model was described in \cite{Caputa:2014eta}.  According to the HRT formula the holographic entanglement entropy (HEE) of the interval $A\in(\ell_1,\ell_2)$ at time $t$ is given by the geodesic $\gamma(x)=\gamma(x,\ell_1,\ell_2,t)$ anchored on the boundary at $x=\ell_1$ and $x=\ell_2$, i.e. satisfying $\gamma(\ell_1,\ell_1,\ell_2,t)=0$ and $\gamma(\ell_2,\ell_1,\ell_2,t)=0$. The HRT formula is
\be 
S_{HEE}(\ell_1,\ell_2,t)=\frac{{\cal L}}{4G},
\ee 
where ${\cal L}={\cal L}(\ell_1,\ell_2,t)$ is the (renormalized) length of $\gamma$.
In general there are two ways how to obtain  ${\cal L}$. The first one is to compute the geodesic length in the background \eqref{metr1}   and then using \eqref{coordtr} map it to the planar BTZ coordinates \eqref{pbtz}. Following this way one can get the exact answer for ${\cal L}(\ell_1,\ell_2,t)$ where $\ell_1$, $\ell_2$ and $t$ are arbitrary.
Another way is to get  good approximation to ${\cal L}$ and to this one  proceeds as follows. First one has to fix $\gamma$ to be that of the unperturbed background  corresponding to $\mu=0$, i.e. we fix $\gamma=\gamma_{BTZ}$ where
\be\label{eq:gamma}
\gamma_{BTZ}(x,\ell_1,\ell_2)=\frac{2e^{\sqrt{M} \left(\ell_1+\ell_2\right)/2} }{\sqrt{M}}\frac{ \sqrt{\sinh
   \left(\sqrt{M} \left(x-\ell_1\right)\right) \sinh \left(\sqrt{M} \left(\ell_2-x\right)\right)}}{e^{\sqrt{M} \ell_1}+e^{\sqrt{M} \ell_2}},
\ee 
is the constant time slice geodesic in the BTZ black hole.
Then we approximate length ${\cal L}$ of this curve by the length of curve given by parametrization \eqref{eq:gamma} (unperturbed geodesic) embedded in the perturbed metric $g$. In another words we compute ${\cal L}$ from the induced metric of unperturbed geodesic in deformed background $g$. In \cite{Nozaki:2013wia,Jahn:2017xsg} it was shown, that this procedure leads to very good qualitative agreement with the exact answer and when $\mu$ is small this agreement become quantitative.

\begin{figure}[h!]
\centering
\includegraphics[width=8.5cm]{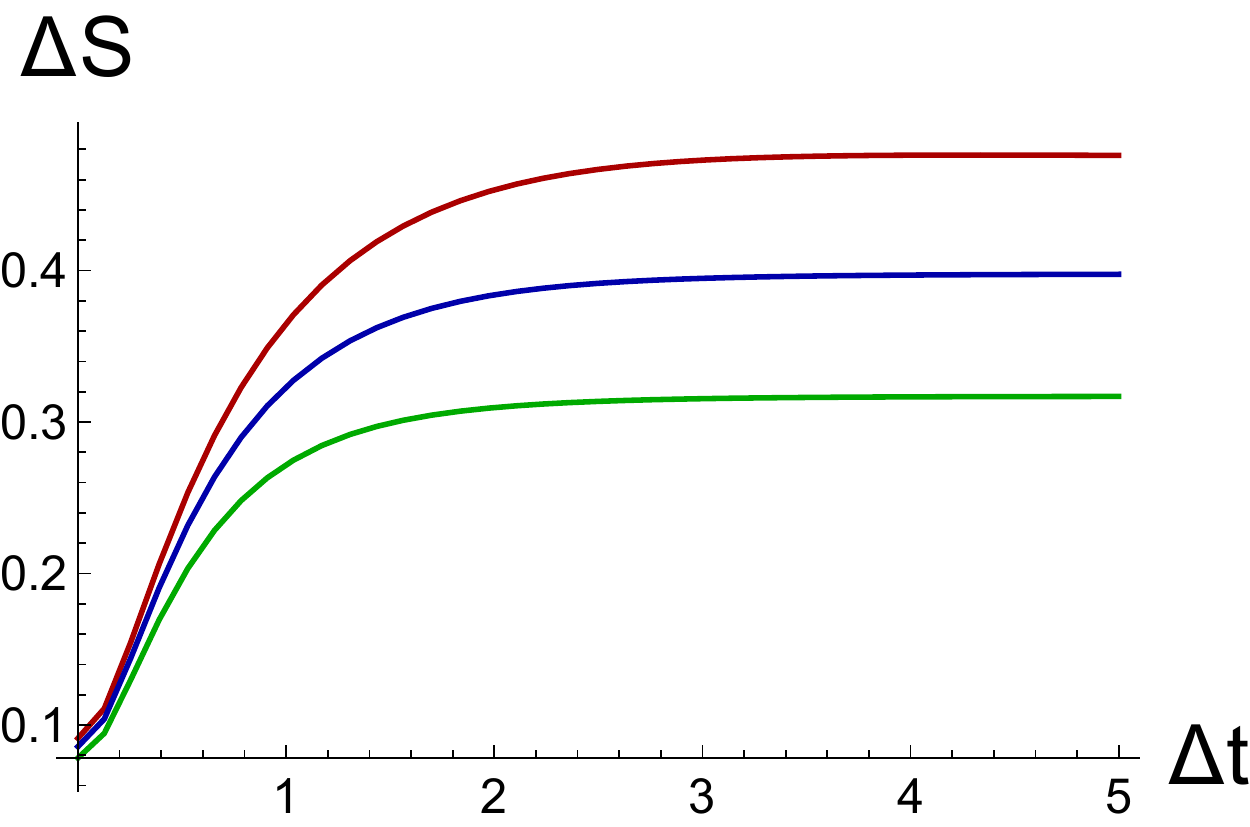}
 \caption{The evolution of the entanglement entropy for the semi-infinite system $x>0$. In this plot $\mu=0.1$, $\varepsilon=0.25$. Curves of different colors correspond to $M=0.5,1,2$ from top to bottom respectively.}
 \label{fig:Sev}
\end{figure}

In Fig.\ref{fig:Sev} we present the evolution of the entanglement entropy of the semi-infinite subsystem $x>0$. In contrast to the  case $T=0$ where the entanglement shows the unbounded logarithmic growth at late time for $T>0$ evolution entanglement saturates to constant value after some time.

\section{Action complexity}\label{sec:CA}

\begin{figure}[h!]
\centering
\includegraphics[width=7.5cm]{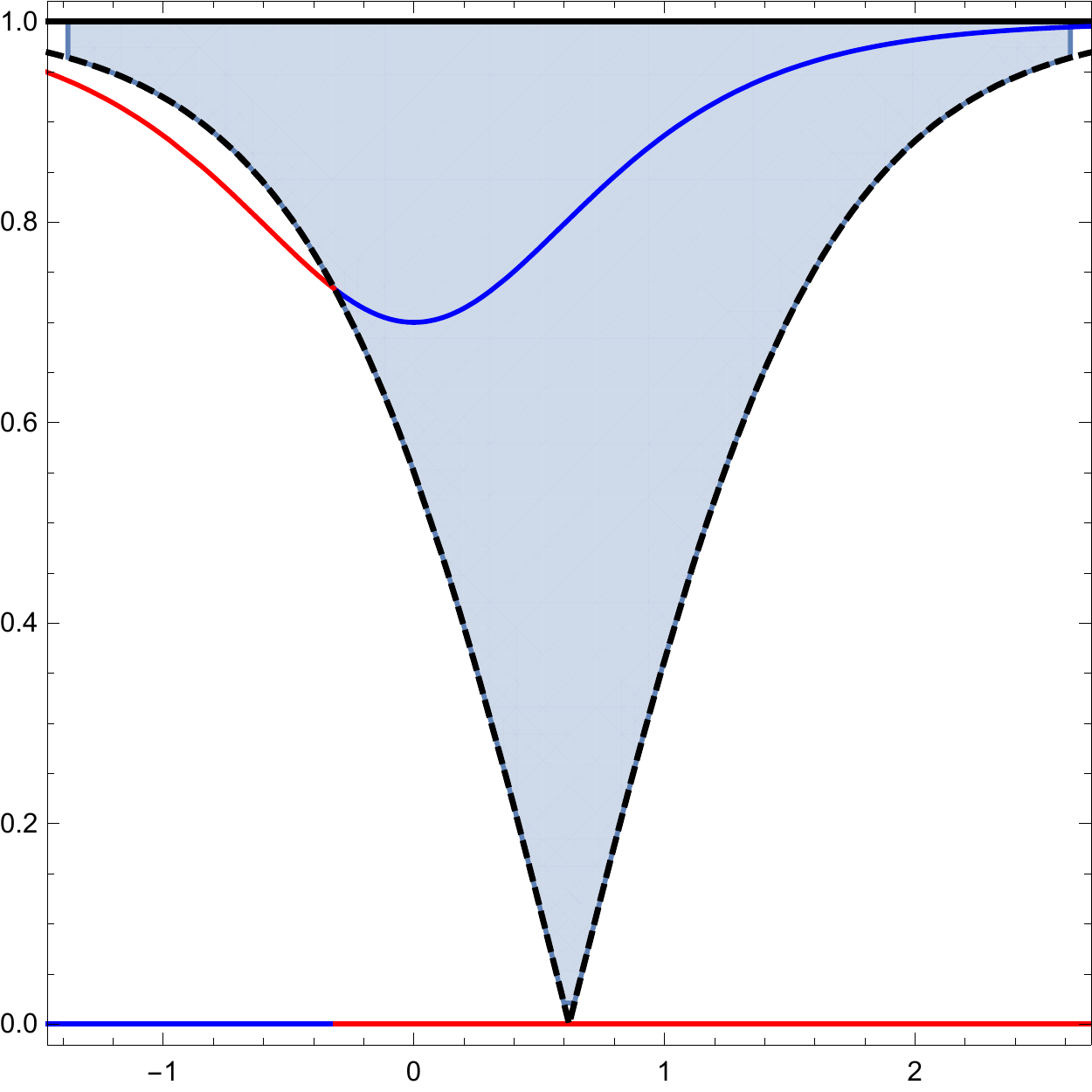}
 \caption{The WdW patch of the state at time moment $\tau=0.62$ corresponds to the blue zone. The contributing part of the worldline is show by the blue curve, red curve corresponds to the part the does not contribute. Here $z_h=1$.}
 \label{fig:CAwdw}
\end{figure}

 Consider now how the CA conjecture works in our setup. For simplicity  consider the probe particle approximation. This means that  we neglect the backreaction of the particle. This implies that the action of our system consists of the gravity action on the static background and the particle action. The model of the particle falling on the black hole in the probe approximation still serves as a nontrivial model dual of precursor perturbation evolving in quantum system at finite temperature (also see the CA probe approximation for the string in $AdS$ in \cite{Nagasaki-1},\cite{Nagasaki-2}). Following the CA prescription the complexity of the state at time $\tau$ is proportional to  the on-shell gravitational and matter action restricted to the special region bounded by null rays emanating from the boundary at time $\tau$. This region is called the Wheeler-de-Witt(WdW) patch  and the action $I_{\text{WdW}}$ in this patch is related to the complexity as
\be 
\Delta {\cC}_{CA}=\frac{I_{\text{WdW}}}{\pi}.
\ee 
In the probe approximation the time-dependent part of the complexity is the particle action swept by the WdW patch.    The probe limit is easily generalized from the $AdS_{3}$ case to general dimension. Null rays bounding the WdW patch corresponding to the state at boundary time $\tau$ has the form
\be
z=\pm z_h\tanh\frac{t-\tau}{z_h}.
\ee 
Massive particle action $S$ between time moments $t_1$ and $t_2$  is given by
\bea\label{eq:CAint}
&&S=-m L\int_{t_1}^{t_2}\frac{2 \bar E}{\left(\bar E^2 z_h^2+1\right) \cosh \left(\frac{2 t}{z_h}\right)-\bar E^2 z_h^2+1}dt,\\
&&\bar{E}=\frac{E}{m L}.
\eea
The intersection of the WdW patch and the particle worldline is presented in Fig.\ref{fig:CAwdw}.
Note, that  from Fig.\ref{fig:CAwdw} one can see that the complexity of the state depends on times $t<0$ (in Fig.\ref{fig:CAwdw} blue part of the worldline curve extends to the region $t<0$).  Remind,  that quench position is $\tau=0$ and $x=0$.  This means, that according to the canonical CA prescription the complexity of the state after quench strongly depends on the preparation protocol of the state. Also this could be the drawback of the particular holographic model of   local quench. Note that this model has time-reversal symmetry (also see the discussion on the related issue in \cite{Ageev-quench,Ageev-quench-2}).
For $\tau>0$ time $t_1$ in the formula \eqref{eq:CAint} is given by
\be
t_1=z_h \log \left(\sqrt{\frac{\left(2 z_h\sqrt{ \left(z_h^2-\varepsilon
   ^2\right) \sinh ^2\left(\frac{\tau }{z_h}\right)}-\varepsilon
   ^2\right) \left(\sinh \left(\frac{\tau }{z_h}\right)+\cosh
   \left(\frac{\tau }{z_h}\right)\right)}{\left(2 z_h^2-\varepsilon
   ^2\right) \sinh \left(\frac{\tau }{z_h}\right)+\varepsilon ^2 \cosh
   \left(\frac{\tau }{z_h}\right)}}\right),
\ee
and $t_2\rightarrow \infty$. After integration the explicit formula for the complexity growth  has the form
\be 
\Delta {\cC}=-\frac{m L}{\pi}\left(\text{arccot}\left(\frac{2 \pi T}{\bar E}\right)+\text{arccot}\left(\frac{2 \pi  T \sinh (2 \pi  T   \tau)}{\sqrt{\bar E^2+4 \pi ^2 T^2}-\bar E \cosh (2 \pi  T \tau)}\right)\right).
\ee 
\begin{figure}[h!]
\centering
\includegraphics[width=7.5cm]{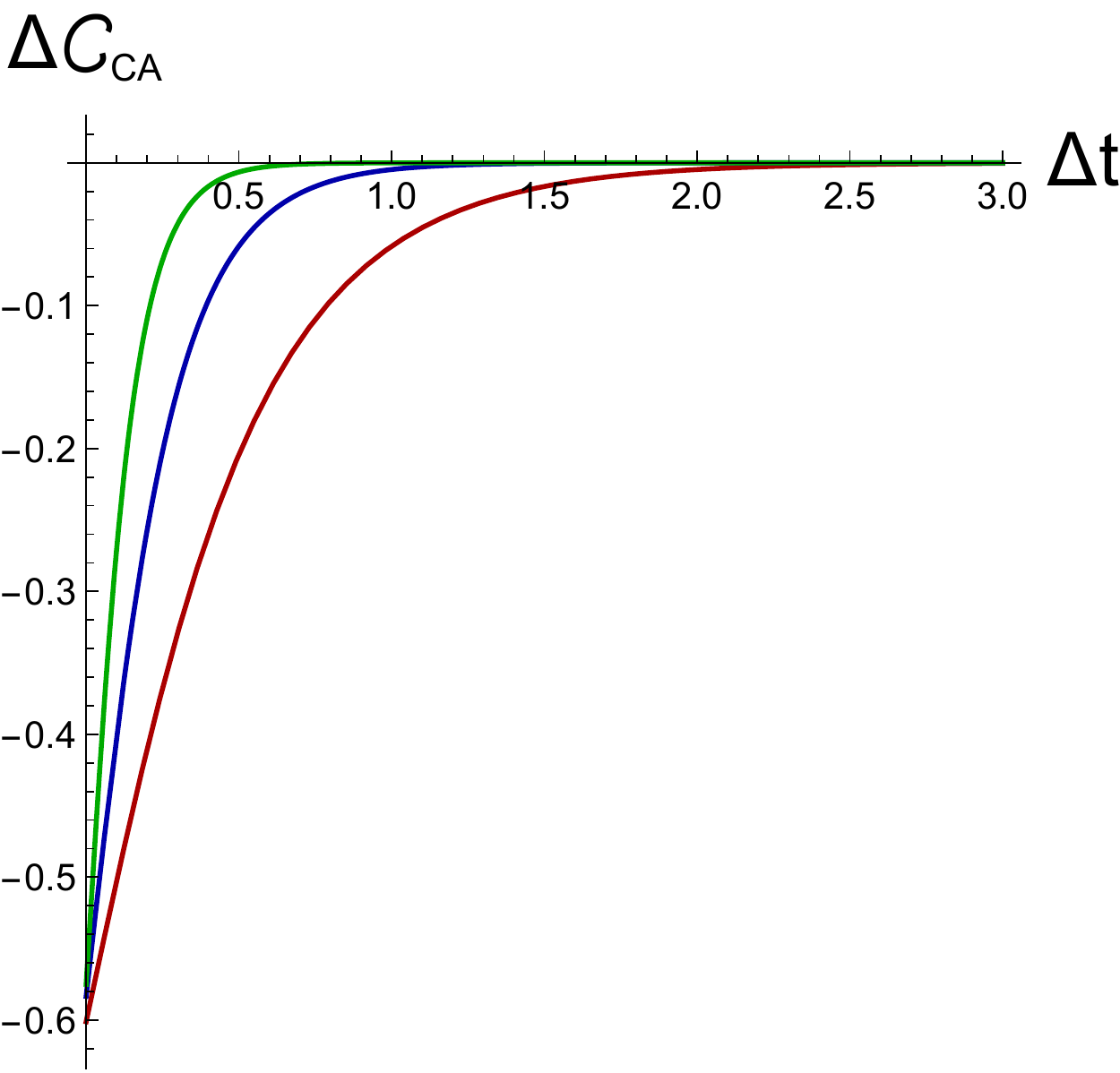}
 \caption{The growth of the CA complexity after a local quench calculated in a probe approximation. Green curve corresponds to the temperature $T=1.5$, blue one to $T=0.8$ and red one to $T=0.8$. The energy of the particle $E=0.25$ is fixed. In this plot we assume that the complexity is normalized, i.e. $\Delta {\cal C}_{CA}=\Delta {\cal C}_{CA}/(mL)$. }
 \label{fig:CAgrowth}
\end{figure}
At the initial time moment $\tau \rightarrow 0$  the complexity exhibits the linear growth of the form
\be 
\Delta {\cC}\approx-h\left(1+\frac{2 h }{\pi }\text{arccot}\left(\frac{4 h   \pi  T}{E}\right)\right)+\frac{\left(E+\sqrt{E^2+16 h^2 \pi ^2   T^2}\right)  }{\pi } \tau,
\ee 
where we used $m=2h/L$.
For the limit $T\rightarrow 0$ we reproduce the result of \cite{Ageev-quench,Ageev-quench-2} where the saturation of Lloyds bound at the initial  time moment $\tau \rightarrow 0$ after perturbation was found. The temperature $T>0$ correction to the complexification rate at $\tau \rightarrow 0$  has the form
\be \label{LloydCA}
\frac{d\Delta {\cC}}{dt}\approx\frac{2 E}{\pi }+\frac{8 h^2 \pi  T^2}{E}.
\ee 
From \eqref{LloydCA} we  see that temperature corrections lead to the violation of the Lloyd bound for all values of temperature. The complexity for subsystems has the qualitative behaviour similar to that one observed in \cite{Ageev-quench,Ageev-quench-2} so we do not analyze it here in details.

\section{Volume complexity}\label{sec:CV}
To calculate  the CV complexity for the stationary dual background one have to find the volume enclosed by the  minimal hypersurface spanned on the boundary at time $\tau$. There are different versions of covariant generalizations of the CV complexity proposal. In this paper we use the approximation similar to that of described in Sect.\ref{sec:ent} for the entanglement entropy. We approximate the volume complexity as
\be \label{eq:CV}
{\cal C}_{CV}=\frac{{\cal V}}{G L},
\ee 
where ${\cal V}$ is the volume of the constant time slice of background with the metric $g$  at the fixed time $\tau$
\be 
{\cal V}=\int\int\left(\sqrt{\det\Sigma_{\tau}}-\frac{1}{z^2\sqrt{1-Mz^2}}\right)dxdz.
\ee 
Here  $\Sigma_\tau$ is the constant time slice volume form and $\Sigma_\tau$ is renormalized  with respect to the constant time slice volume of the BTZ black hole ($\mu=0$).

$\,$

\begin{figure}[h!]
\centering
\includegraphics[width=5.45cm]{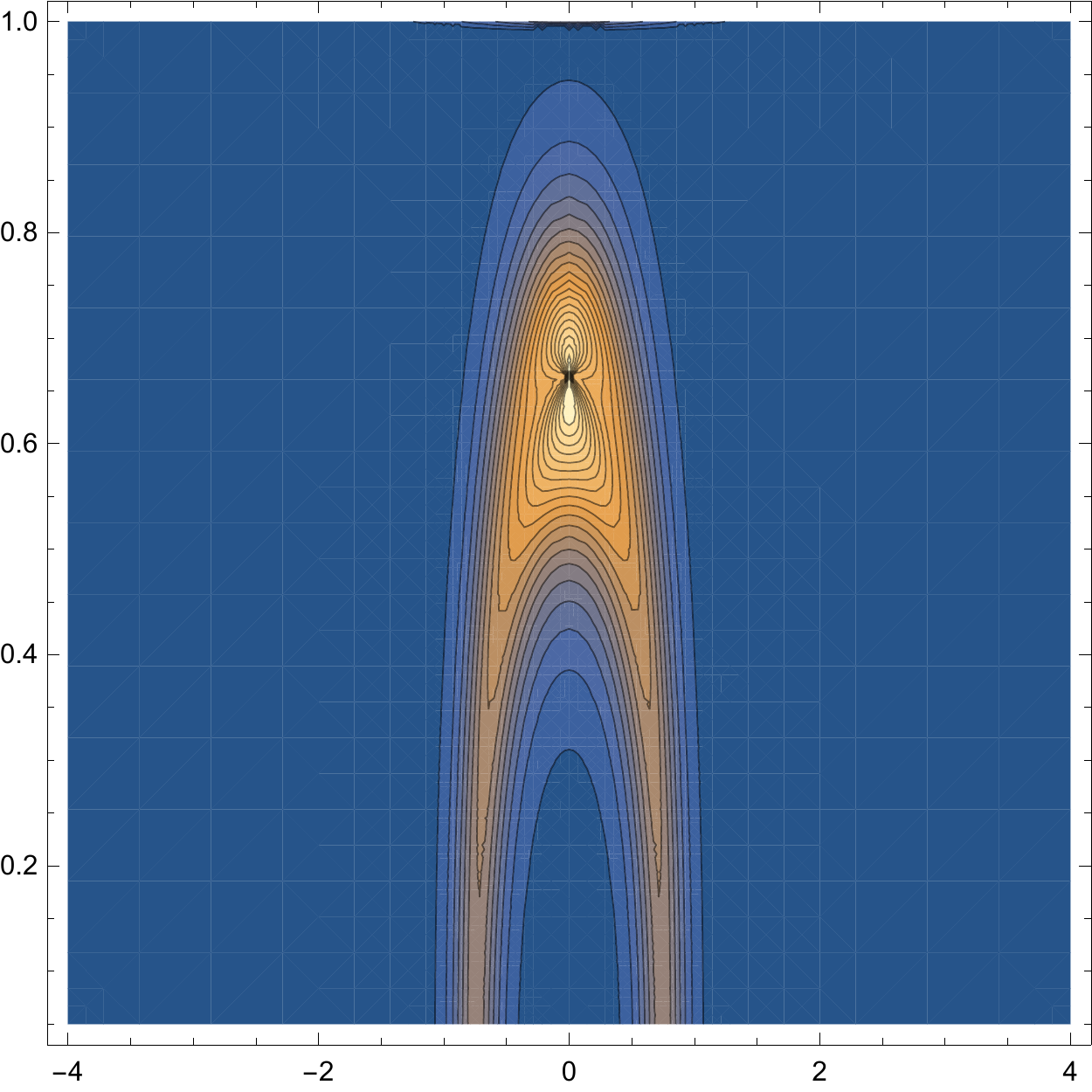}
\includegraphics[width=6.5cm]{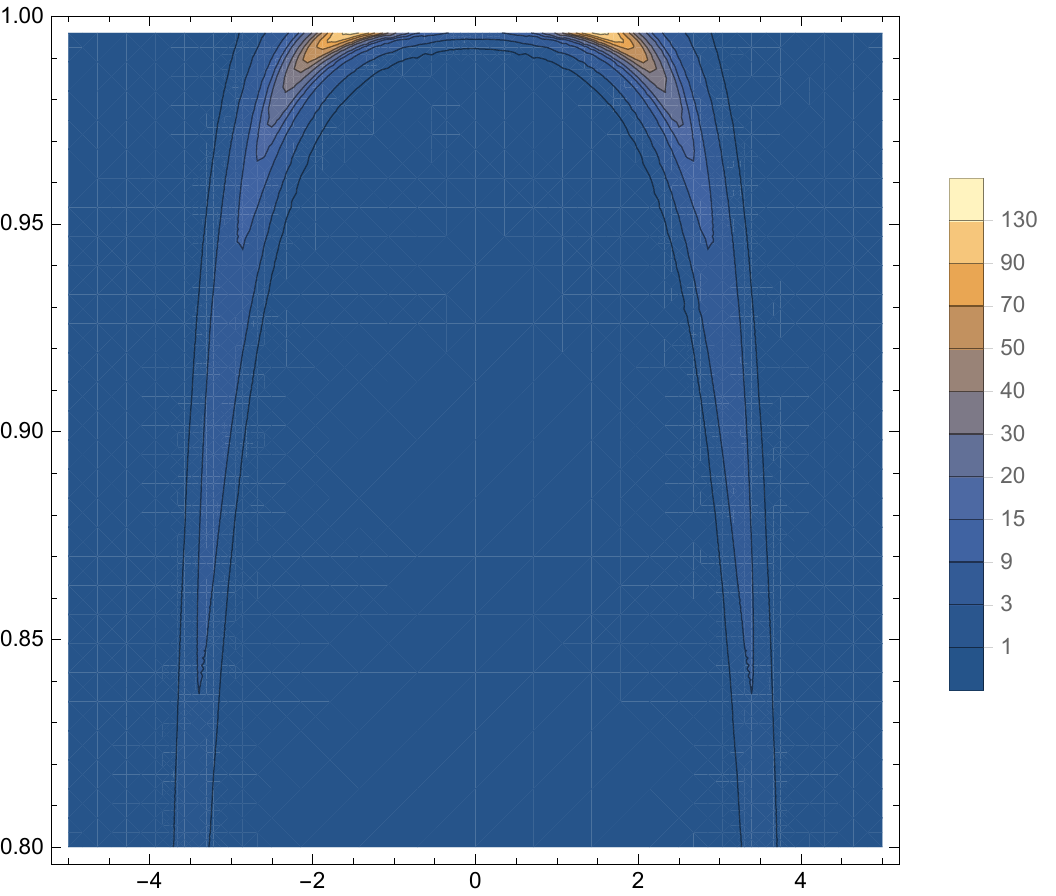}
 \caption{The density of the (renormalized) volume $\Sigma$ with $\mu=0.1$, $M=1$, $\varepsilon=0.25$. The left plot corresponds to $\tau=0.75$ and the right one to $\tau=4$.}
 \label{fig:contour}
\end{figure}

To get some qualitative understanding of the gravitational dynamics of this model we present plot of $\Sigma_{\tau}$ in Fig.\ref{fig:contour}. We see that at the inital time moment $\tau=0$ gravitational perturbation starts from the boundary $z=0$. Then the perturbation propagates to the horizon. Near the horizon at some time it splits into two parts consisting of two large tails touching the boundary approximately at $\tau=x$. Each tail become more dense closer to horizon. In the near-horizon zone they form kind of very dense gravitational perturbation. The whole late-time structure looks like "wormhole" connecting position of two "quasiparticles" in the boundary theory.

\begin{figure}[h!]
\centering
\includegraphics[width=9.5cm]{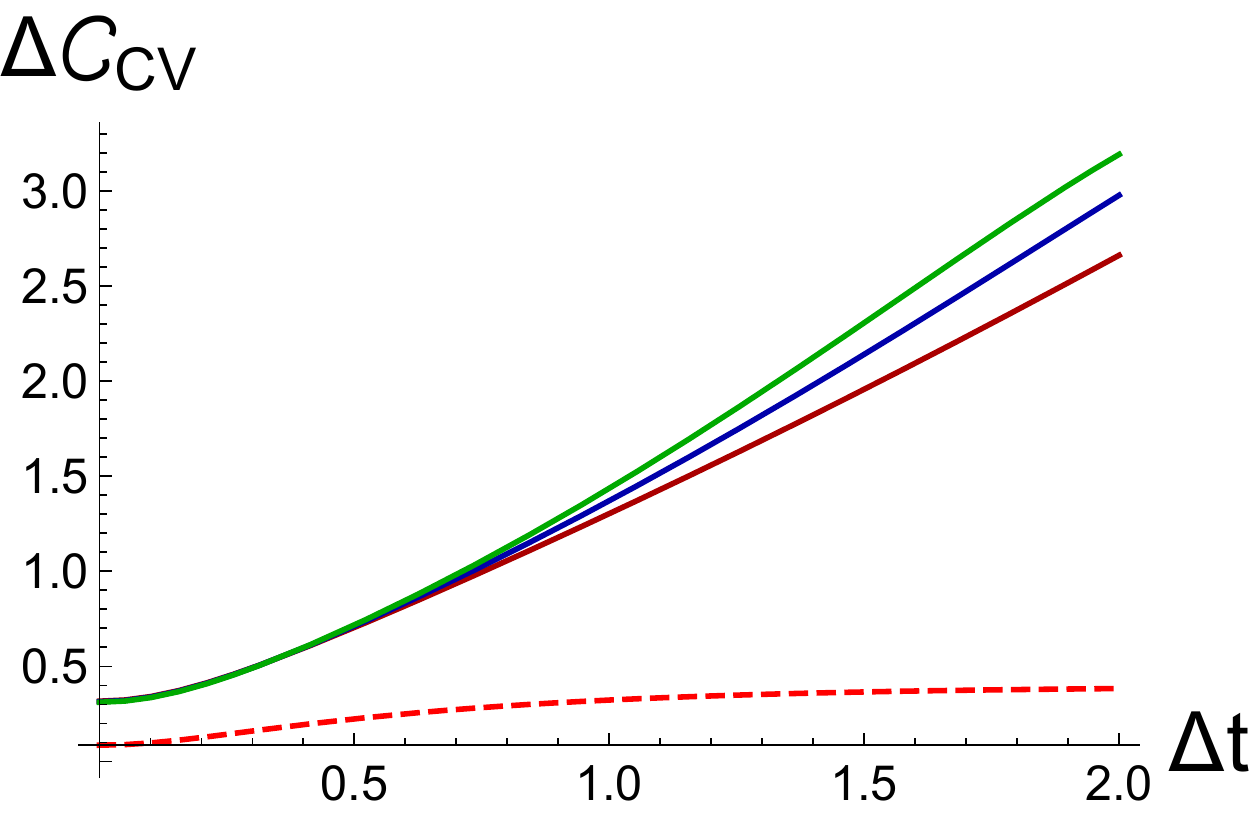}
 \caption{The CV complexity evolution of the total system for different values of temperature. Different solid curves correspond to different colors: $M=1$ (red curve), $M=1.5$ (blue curve) and $M=2$ (green curve) with the parameter values $\varepsilon=0.25$ and $\mu=0.1$. Dashed line corresponds to the entanglement evolution for the semi-infinite system $x>0$ with $M=1$. Note, that here we plot the rescaled CV complexity, i.e. ${\cal C}_{CV}\rightarrow{\cal C}_{CV} \cdot G L$}
 \label{fig:cnt}
\end{figure}

Now let us turn to the description of the complexity growth of the total state using formula \eqref{eq:CV}. Calculating this integral numerically  we get the time dependence with fixed $\varepsilon$, $\mu$ and $M$. We present the result of the calculation in Fig.\ref{fig:cnt}. We see, that at the initial  stage $\tau\rightarrow 0$ we get the quadratic growth. 
This is the common point when the dual description involves geometric quantities like the hypersurfaces stretched from the boundary. After the quadratic  growth there is the regime of linear unbounded growth of the complexity $\Delta{\cal C}$.  In the case of $T=0$ we see that  this growth  is not linear.
The late time linear growth is the consequence of the fast unbounded growth of the gravitational perturbation that spreads over the black hole background. 

Let us outline now how this is reflected in the behaviour of the subsystem CV complexity. First let us consider the single  interval as the subsystem.  The evolution is  similar to the total system case except times $t\approx\ell$ where one can observe sharp decay of complexity to the equilibrium values.   When $\ell \rightarrow \infty$ the point of decay is moving to $t\rightarrow\infty$ and we get an infinite linear growth corresponding to the linear growth of the total system. By naive dimensional analysis one can estimate this  universal linear growth (of volume) as 
\be 
\Delta {\cal V} \sim \frac{\mu M}{\varepsilon} \cdot t
\ee 
and the numerical calculations with small $\varepsilon$ and $\mu$ shows that this relation takes place. Using this result one  derive the linear growth of complexity in the form
\be 
\Delta {\cal C} \sim 32 \pi ^2 T^2 \sqrt{E^2+\pi ^2 h^2 T^2} \cdot t
\ee
which violates the Lloyds bound for some values of $T$ and $h$.

\begin{figure}[h!]
\centering
\includegraphics[width=9.5cm]{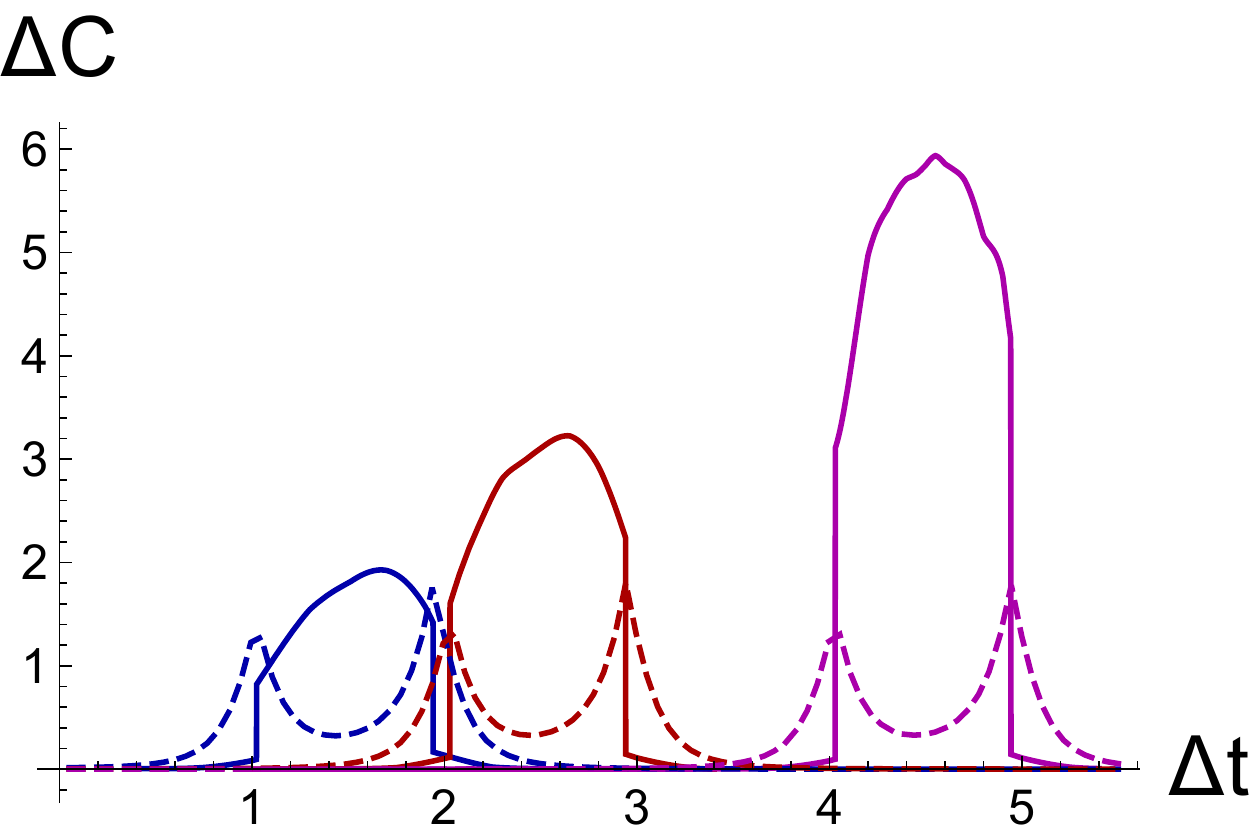}
 \caption{The CV complexity evolution(solid curves) and the entanglement entropy(dashed curve) for  two disjoint intervals subsystem. The blue curves correspond to subsystem $x\in(-2,1)\cup(1,2)$, red curves to $x\in (-3,-2)\cup(2,3)$ and the magenta ones to $x\in(-5,-4)\cup(4,5)$. The entanglement entropy values are rescaled in 20 times to compare with the complexity and the complexity is rescaled as ${\cal C}_{CV}\rightarrow{\cal C}_{CV} \cdot G L$.}
 \label{fig:CC2}
\end{figure}

Now let us consider the complexity evolution for two disjoint intervals. In this case the comparison of the complexity of this subsystem  with the entanglement evolution shows the difference between these two quantities. For simplicity we take two intervals of the length  $\ell$ separated by the distance $d$ located symmetric with respect to the quench point $x=0$. We present the evolution of entanglement and complexity for this system in Fig.\ref{fig:CC2}.
The entanglement evolution shows two sharp peaks approximately at times $\tau_1=d/2$ and $\tau_2=d/2+\ell$. Between these two times the entanglement has have minimum approximately at $\tau_3=d/2+\ell/2$. There are two sets of competing  HRT surfaces connecting intervals - the "disjoint" set corresponding to the geodesics spanned on the  interval on each side of quench separately  and the "connected" set corresponding to the geodesics spanned on the endpoints of different intervals. On the early and late (before the first peak and after the last peak in the entanglement) stages of evolution the disconnected set dominates. This leads to the slow growth and decay of complexity at these stages. However in the intermediate regime between $\tau_1$ and $\tau_2$ the complexity exhibits the discontinous jump, then fast saturation  to the maximum and rapid decay approximately at time corresponding to entanglement minimum. For  $d\rightarrow\infty$ the maximal value of complexity is growing (in contrast to the $T=0$ where this growth is very slow for large $d$).

\newpage

\section{Discussion and summary}\label{sec:disc}
This paper is devoted to the study of the holographic complexity of a finite temperature state in two-dimensional conformal field theory that evolves after the perturbation by the local precursor. This precursor is  the primary operator in the CFT with the scaling dimension $h$. From the holographic viewpoint this process is described  by the point particle falling on the one-sided planar BTZ black hole horizon. The exact metric for this process is known explicitly. We calculate the CA and the CV complexity evolution in this holographic background. We make the approximate calculation for both CA and CV prescriptions. For the CA proposal we restrict ourselves to the case of probe particle approximation. We find the explicit formula describing the evolution of the action complexity. The CA evolution consists  of the linear growth stage and the saturation at the late times. Instead of the saturation one can expects expect here the unbounded linear growth as different works predict \cite{Susskind:2018pmk}, \cite{Stanford:2014jda}. To get the late time linear growth one has to involve additional assumptions.   There are different examples existing in the literature where one has to involve additional considerations to get the desired linear growth of the system. These examples are different dual models of the one-dimensional strongly coupled systems \cite{Brown:2018bms}. In \cite{Brown:2018bms} one introduces the special boundary terms in the action and in \cite{Alishahiha:2018swh} one uses the finite bulk cutoff to get linear growth. It would be interesting to investigate the modification of our setup to get the linear late time growth. Also this calculation is performed in the  approximate regime. So taking into account the gravitational backreaction effects may change the situation and produce the desired linear growth behaviour from the first principles. 

As in thermofield state the CV conjecture calculation demonstrates the late time unbounded linear growth without any additional considerations. However in contrast to the CA conjecture the CV complexity shows the initial quadratic growth usually corresponding to the equilibration of the local correlations in the system. We give the simple estimate of the linear growth coefficient and find, that it violates Lloyds bound for some temperature values. The intuitive picture of dynamics of geometrical quantities in this model is as follows.  After the quench gravitational perturbation evolves and get to the black hole horizon position. After that the perturbation splits in two localized "tubes" connected through the near-horizon region. Of course this calculation is performed in the constant time approximation, so there is a question of the regime of validity of this approximation. The picture shown here also supports the idea, that the gravitational contribution may change the CA evolution in the desired way. Finally we compare the entanglement and complexity evolution for  system of disjoint intervals and find, that the (local) minimum of the entanglement is realized when the complexity is maximal. This raises  question about how the complexity is related to the quasiparticle picture. Remember, that the entanglement is in good correspondence with this picture. The further investigations and possible extensions of this work include the comparison of the TFD state perturbed by local operator with one-sided black hole. Also it is important to consider the inclusion of the gravitational correction to the CA computation. Also it is interesting to compare with different complexity results concerning non-equilibrium states developed in \cite{QFT1}-\cite{QFT2}.

\section*{Acknowledgements}
I would like to thank I.Ya.Aref'eva for comments on the early version of this text. This work is supported by the Russian Science Foundation (project 17-71-20154)

\end{document}